\documentstyle[epsf]{mn}
\newif\ifAMStwofonts
\AMStwofontstrue

\makeatletter
\def\figsize{\ifSFB@referee\epsfxsize=0.5\hsize\else\epsfxsize=\hsize\fi}
\newif\ifdraft \def\draft{\ifSFB@referee\drafttrue\fi}
\makeatother
\def\eq#1#2 {\begin{equation} \if!#2!{#1}\else\label{#1}#2\fi \end{equation}
      \ifdraft\if!#2!\else\marginpar{\small #1}\fi\fi}
\def\eqarray#1#2 {\begin{eqnarray}\if!#2!{#1}\else\label{#1}#2\fi\end{eqnarray}
      \ifdraft\if!#2!\else\marginpar{\small #1}\fi\fi}

\def\Ref{\bibitem{}}

\def\r    {\hbox{\bf r}}
\def\y    {\hbox{\bf y}}

\def\mic  {\hbox{$\umu$m}}
\def\HII  {\hbox{H\small II}}

\def\symbol#1{\hbox{$#1$}}
\def\sub#1{\hbox{$_{\rm #1}$}}
\def\about{\symbol{\sim}}
\def\x    {\symbol{\times}}
\def\B    {\symbol{B_\lambda}}
\def\I    {\symbol{I_\lambda}}
\def\F    {\symbol{F_\lambda}}
\def\J    {\symbol{J_\lambda}}
\def\Fd   {\symbol{F_{\rm d\lambda}}}
\def\Jd   {\symbol{J_{\rm d\lambda}}}
\def\S    {\symbol{S_\lambda}}
\def\b    {\symbol{b_\lambda}}
\def\u    {\symbol{u_\lambda}}
\def\f    {\symbol{f_\lambda}}
\def\k    {\symbol{\kappa_\lambda}}
\def\ka   {\symbol{\kappa_{\rm a\lambda}}}
\def\ks   {\symbol{\kappa_{\rm s\lambda}}}
\def\q    {\symbol{q_{\lambda}}}
\def\qa   {\symbol{q_{\rm a\lambda}}}
\def\qP   {\symbol{q_{\rm P}}}
\def\qaP  {\symbol{q_{\rm aP}}}

\def\qae  {\symbol{q_{\rm ae}}}
\def\qF   {\symbol{q_{\rm F}}}
\def\t    {\symbol{\tau_\lambda}}
\def\tT   {\symbol{\tau_\lambda^T}}
\def\tV   {\symbol{\tau_{\rm V}}}
\def\al   {\symbol{\varpi_\lambda}}
\def\P    {\symbol{\cal P}}
\def\Le   {\symbol{L_{\rm e}}}
\def\Te   {\symbol{T_{\rm e}}}
\def\Ie   {\symbol{I_{\rm e\lambda}}}
\def\fe   {\symbol{f_{\rm e\lambda}}}
\def\re   {\symbol{r_{\rm e}}}
\def\rre  {\symbol{r_{\rm e}^2}}

\def\yout {\symbol{y_{\rm out}}}
\def\Fone {\symbol{F_{\rm e1}}}
\def\Tsub {\symbol{T_{\rm sub}}}
\def\Teff {\symbol{T_{\rm eff}}}
\def\lp   {\symbol{\lambda_{\rm p}}}
\def\lout {\symbol{\lambda_{\rm out}}}

\ifoldfss
  \ifCUPmtlplainloaded \else
    \NewTextAlphabet{textbfit} {cmbxti10} {}
    \NewTextAlphabet{textbfss} {cmssbx10} {}
    \NewMathAlphabet{mathbfit} {cmbxti10} {} % for math mode
    \NewMathAlphabet{mathbfss} {cmssbx10} {} %  "   "    "
  \fi
  \ifAMStwofonts
    \ifCUPmtlplainloaded \else
      \NewSymbolFont{upmath} {eurm10}
      \NewSymbolFont{AMSa} {msam10}
      \NewMathSymbol{\upi}     {0}{upmath}{19}
      \NewMathSymbol{\umu}     {0}{upmath}{16}
      \NewMathSymbol{\upartial}{0}{upmath}{40}
      \NewMathSymbol{\leqslant}{3}{AMSa}{36}
      \NewMathSymbol{\geqslant}{3}{AMSa}{3E}

       \let\le=\leqslant
       
    \fi
  \fi
\fi % End of OFSS

\ifnfssone
  \newmathalphabet{\mathit}
  \addtoversion{normal}{\mathit}{cmr}{m}{it}
  \addtoversion{bold}{\mathit}{cmr}{bx}{it}
  \newmathalphabet{\mathbfit} % math mode version of \textbfit{..}
  \addtoversion{normal}{\mathbfit}{cmr}{bx}{it}
  \addtoversion{bold}{\mathbfit}{cmr}{bx}{it}
  \newmathalphabet{\mathbfss} % math mode version of \textbfss{..}
  \addtoversion{normal}{\mathbfss}{cmss}{bx}{n}
  \addtoversion{bold}{\mathbfss}{cmss}{bx}{n}
  \ifAMStwofonts
    \ifCUPmtlplainloaded \else
      \UseAMStwoboldmath
      \makeatletter
      \new@mathgroup\upmath@group
      \define@mathgroup\mv@normal\upmath@group{eur}{m}{n}
      \define@mathgroup\mv@bold\upmath@group{eur}{b}{n}
      \edef\UPM{\hexnumber\upmath@group}
      \new@mathgroup\amsa@group
      \define@mathgroup\mv@normal\amsa@group{msa}{m}{n}
      \define@mathgroup\mv@bold\amsa@group{msa}{m}{n}
      \edef\AMSa{\hexnumber\amsa@group}
      \makeatother
      \mathchardef\upi="0\UPM19
      \mathchardef\umu="0\UPM16
      \mathchardef\upartial="0\UPM40
      \mathchardef\leqslant="3\AMSa36
      \mathchardef\geqslant="3\AMSa3E

       \let\le=\leqslant

    \fi
  \fi
\fi % End of NFSS release 1

\ifnfsstwo
  \DeclareMathAlphabet{\mathbfit}{OT1}{cmr}{bx}{it}
  \SetMathAlphabet\mathbfit{bold}{OT1}{cmr}{bx}{it}
  \DeclareMathAlphabet{\mathbfss}{OT1}{cmss}{bx}{n}
  \SetMathAlphabet\mathbfss{bold}{OT1}{cmss}{bx}{n}
  \ifAMStwofonts
    \ifCUPmtlplainloaded \else
      \DeclareSymbolFont{UPM}{U}{eur}{m}{n}
      \SetSymbolFont{UPM}{bold}{U}{eur}{b}{n}
      \DeclareSymbolFont{AMSa}{U}{msa}{m}{n}
      \DeclareMathSymbol{\upi}{0}{UPM}{"19}
      \DeclareMathSymbol{\umu}{0}{UPM}{"16}
      \DeclareMathSymbol{\upartial}{0}{UPM}{"40}
      \DeclareMathSymbol{\leqslant}{3}{AMSa}{"36}
      \DeclareMathSymbol{\geqslant}{3}{AMSa}{"3E}

       \let\le=\leqslant

    \fi
  \fi
\fi % End of NFSS release 2

\ifCUPmtlplainloaded \else
  \ifAMStwofonts \else % If no AMS fonts
    \def\upi{\pi}
    \def\umu{\mu}
    \def\upartial{\partial}
  \fi
\fi

\pagerange{\pageref{firstpage}--\pageref{lastpage}}
\pubyear{1997}

\title[Self-similarity and scaling of dust IR emission]
       {Self-similarity and scaling behavior of IR emission
                  from radiatively heated dust: I. Theory}

\author[\v{Z}eljko Ivezi\'{c} and Moshe Elitzur]
                     { \v{Z}eljko Ivezi\'{c} and Moshe Elitzur \\
                        Department of Physics and Astronomy,
                 University of Kentucky, Lexington, KY 40506-0055, USA\\
                   e-mail: ivezic@pa.uky.edu, moshe@pa.uky.edu }

\date{Accepted December 16, 1996. Received September 30, 1996; in original form July 17, 1996}
\begin{document}
\maketitle

\label{firstpage}

\begin                             {abstract}

Dust infrared emission possesses scaling properties that yield powerful results  
with far reaching observational consequences.  Scaling was first noticed by  
Rowan-Robinson for spherical shells and is shown here to be a general property  
of dust emission in arbitrary geometries. Overall luminosity is never an input 
parameter of the radiative transfer problem, spectral shape is the only relevant 
property of the heating radiation when the inner boundary of the dusty region is 
controlled by dust sublimation.  Similarly, the absolute scales of densities and 
distances are irrelevant; the geometry enters only through angles, relative 
thicknesses and aspect ratios, and the actual magnitudes of densities and 
distances enter only through one independent parameter, the overall optical 
depth.  That is, as long as the overall optical depth stays the same, the system 
dimensions can be scaled up or down by an arbitrary factor without any effect on 
the radiative transfer problem. Dust properties enter only through 
dimensionless, normalized distributions that describe the spatial variation of 
density and the wavelength dependence of scattering and absorption efficiencies.

Scaling enables a systematic approach to modeling and classification of IR
spectra. We develop a new, fully scale-free method for solving radiative
transfer, present exact numerical results, and derive approximate analytical
solutions for spherical geometry, covering the entire range of parameter space
relevant to observations.  For a given type of grains, the spectral energy
distribution (SED) is primarily controlled by the profile of the spatial dust
distribution and the optical depth --- each density profile produces a family
of solutions, with position within the family determined by optical depth.
From the model SEDs presented here, the density distribution and optical depth
can be observationally determined for various sources.

Scaling implies tight correlations among the SEDs of various members of the
same class of sources such as young stellar objects, late-type stars, etc.  In
particular, all members of the same class occupy common, well defined regions
in color-color diagrams.  The observational data corroborate the existence of
these correlations.

\end{abstract}


\section                         {INTRODUCTION}

For many astronomical objects, the radiation we receive has undergone
significant processing by surrounding dust, and its detailed interpretation
requires considerable theoretical effort.  Input to the necessary radiative
transfer calculations includes properties of the original radiation, optical
characteristics of individual dust grains, and the dust density distribution.
Specifying all the relevant properties traditionally involved a rather large
number of input parameters, creating two major practical problems.  First, the
volume of parameter space that must be searched to fit a given set of
observations can become prohibitively large. Second, and more serious, even
when a successful fit is accomplished, its uniqueness is questionable and the
model parameters cannot be trusted as a reliable indication of the source
actual properties.

In fact, much of this input is redundant because the radiative transfer
problem possesses general scaling properties that drastically reduce the
number of independent input parameters.  Rowan-Robinson (1980, RR hereafter)
was the first to utilize scaling in his extensive study of IR emission from
spherical dust shells.  In his formulation of the radiative transfer problem,
the overall luminosity, as well as other quantities, never enters.  The full
power of this scaling was exploited in our previous work on late-type stars
(Ivezi\' c \& Elitzur 1995, IE95 hereafter). There we show that, given the
dust optical properties, for each circumstellar shell both the dynamics and IR
emission are successfully described by a single parameter -- the overall
optical depth. This explains the many correlations displayed by the data for
late-type stars.  Here we derive the general form of these results, extending
the scaling analysis to arbitrary geometries and density distributions. Formal
discussion of the general scaling properties of the radiative transfer
equation is presented in Section 2, and applied in Section 3 to radiatively
heated dust in arbitrary geometries. In Section 4, more detailed, concrete
results are derived for spherically symmetric systems, leading to a new, exact
method for solution of the radiative transfer problem that can be readily
extended to arbitrary geometries.  Thanks to scaling, the effects of the
independent input properties on emerging radiation can be searched
systematically.  In Section 5 we present such an exhaustive search for
spherical systems and the entire relevant range of input properties.  We
explore the effects of various density distributions, grain size and chemical
composition, sublimation temperature, and spectral shape of the external
radiation. Section 6 contains a summary of our results and a discussion of
observational implications.  A subsequent paper will present detailed analysis
of the IRAS data base, which displays strong evidence for the scaling
properties derived here.

\section                   {GENERAL SCALING PROPERTIES}

The equation of radiative transfer in steady-state is
\eq{rad-tran1}{
             {d\I \over d\ell} = \kappa_\lambda(\S - \I).
}
Here $\kappa_\lambda$ is the overall extinction coefficient at wavelength
$\lambda$ including both absorption and scattering, $\k = \ka + \ks$, and the
corresponding optical depth element along the path is $d\t = \k d\ell$.  We
will also make use of the albedo $\al = \ks/\k$.  In addition to boundary
conditions, solution of this equation for the intensity \I\ requires two input
functions at every point \r\ -- the source function \S(\r) and the absorption
coefficient \k(\r).  The equation involves quantities with dimensions of either
intensity or length.  The intensity \I\ and source function \S\ belong to the
first type, the path-length $\ell$ and absorption coefficient \k, which has
dimensions of inverse length, to the second.  Therefore the formulation of any
radiative transfer problem can involve only two dimensional scales, one for
intensities the other for lengths.  All other relevant scales must be
dimensionless, involving ratios with the two dimensional scales.

\subsection {Length Scales}

Consider first the length-type quantities.  Because all distances can be
replaced by equivalent dimensionless \t, the radiative transfer problem is
independent of linear size scales.  The only geometrical properties that can
enter the problem involve angular sizes and the spatial variation of optical
depth.  To further explore this scaling, introduce some arbitrary length scale
$r_1$.  Any position $\r = (r,\theta,\phi)$ can be specified instead by the
dimensionless vector $\y = (y,\theta,\phi)$, where
\eq{rscaling}{
                              y = r/r_1
}
is dimensionless distance from the (arbitrary) origin.  Denote a radiative
transfer path by \P\ and positions along this path by $(y;\P)$, where $yr_1$
is distance from the path's closest approach to the origin. Then the
path-length element is $d\ell= r_1dy$ and the radiative transfer equation
(\ref{rad-tran1}) becomes
\eq{rad-tran2}{
           {d\I(y;\P) \over dy} = \tT(\P)\eta(y;\P)[\S(y;\P) - \I(y;\P)].
}
Here \tT(\P) is the total optical depth along the path at wavelength $\lambda$
and
\eq{eta gen}{
\eta(y;\P) = {1\over\tT(\P)}{d\t(y;\P)\over dy}= {\k(y;\P)\over\int\k(y;\P)dy}
}
is a dimensionless function that describes the spatial variation of opacity. In
general $\eta(y;\P)$ depends also on wavelength. This dependence disappears
when the properties of individual absorbers do not vary with position, in which
case $\eta(y) = n\sub a(y)/\!\!\int n\sub a(y)dy$, the normalized dimensionless
density distribution of absorbers. If in addition the absorber abundance does
not vary, i.e., $n\sub a/n$ is constant, $\eta$ becomes simply the normalized
dimensionless density distribution $n(y)/\!\!\int n(y)dy$.

In its new form, the radiative transfer equation (\ref{rad-tran2}) no longer
contains any quantities with dimensions of length; physical dimensions enter
only as ratios, i.e., angles and aspect ratios.  Furthermore, the length scale
$r_1$ has disappeared altogether, thus it is not required for a complete
solution. This quantity enters into the problem only through the product $r_1\k
= \tT\eta$. Therefore, any scale associated with it is contained in the overall
optical depth \tT\ because the function $\eta$ sets its own scale from the
normalization
\eq{
                           \int\eta(y;\P) dy = 1
}
for any path \P. This shows that {\em radiative transfer is scale invariant}.
The physical dimensions of any system can be scaled up and down arbitrarily
without any effect on its radiative properties as long as optical depths and
relative matter distribution remain the same. Two systems with entirely
different dimensions and absorption coefficients but the same overall optical
depths will produce the same radiative intensities if they have self-similar
distributions of opacities and source functions. Since the length scale $r_1$
does not enter into the radiative transfer problem, a corollary is that it can
never be determined from the solution. The only way to determine actual length
scales is through some independent determination of the size of the system or
the distance to it.

The scale-invariance of radiative transfer reflects the fact that, in principle, 
the problem could be formulated using optical depth as the independent variable, 
differentiating with respect to $d\t(y;\P) = \tT(\P)\eta(y;\P)dy$.  In that 
case, location in space (the argument of \I\ and \S) is specified by its optical 
depth along the chosen path, $\t(y;\P) = \tT(\P) \int^y\!\eta(u;\P)du$. This 
scale-invariant formulation is completely general and applies to all geometries, 
including irregular shapes, clumpy media, etc.  In this general form, the 
path-dependent function $\eta(y;\P)$ describes the geometrical morphology, in 
essence defining the transformation from real space to $\tau$-space, which might 
be quite involved.  In cases of geometrical symmetry, radiative transfer is more conveniently formulated with $\tau$ defined along the axis of symmetry.  Then $\eta(y)$ is defined in an analogous manner, for example along the radial direction in the case of spherical symmetry (cf Section \ref{spherical}).

\section                         {HEATED DUST}

The discussion so far was general and applies to all radiative problems without
restrictions.  In practice, the scale invariance is mostly useful when the
absorption coefficient is independent of intensity, and from here on we
concentrate on such systems. Thus the following discussion excludes line
emission and photoionization (where the absorption coefficient may depend on
intensity through its effect on level populations), but is fully applicable to
continuum radiation -- especially by dust, the main thrust of this paper.  With
this restriction, in the absence of a source function radiative transfer would
be described by a homogeneous equation. That is, if \I\ is a solution when the
source term is removed then so is $a\I$ for any arbitrary constant $a$. This
scale invariance is broken by the source term, which makes the radiative
transfer equation inhomogeneous.  In the case of a dusty medium, the source
function is
\eq{source}{
 \S = (1 - \al)\B(T) + \al\int\I(\Omega')g(\Omega',\Omega){d\Omega'\over4\upi},
}
where \B\ is the Planck function and $g(\Omega',\Omega)$ is the angular phase
function for coherent scattering from direction $\Omega'$ to $\Omega$ (e.g.\
Mihalas 1978).  Because of scattering, in general \S\ contains a dependence on
directions. We assumed for simplicity single-type dust grains, characterized by
a single temperature $T$.  The general grain mixture case is similar in all
fundamental aspects and is discussed separately in appendix A.

Since the scattering term is linear in \I, it preserves the scale invariance of
intensity.  The only term to break this invariance and set a scale for the
intensity is the emission term, $\B(T)$. It does that by introducing the dust
temperature, determined from radiative equilibrium
\eq{
         \int d\Omega\int\k\S d\lambda = 4\upi\int\k\J d\lambda\,,
}
where $\J = \int\I d\Omega/4\upi$ is the angle-averaged intensity.  Both sides
of this relation vary linearly with the extinction coefficient, therefore the
actual magnitude of this quantity is irrelevant -- only its wavelength
variation matters.  Indeed, Leung (1976) noticed from his numerical calculations 
that $T$ was ``rather insensitive" to the absolute values of extinction 
coefficient; our discussion shows that this insensitivity is in fact an exact 
symmetry property of the problem.  To emphasize this point we write this 
relation in terms of
\eq{
                           \q = {\k \over \k_0},
}
the extinction coefficient normalized to unity at some fiducial, arbitrary 
wavelength $\lambda_0$. \footnote{Although the absolute value of \k\ does not 
enter independently into the radiative transfer problem, it is needed to relate 
the optical depth to the total dust mass.}  With the dust source function from 
equation \ref{source}, the radiative equilibrium condition becomes
\eq{equilib}{
                  \int\qa\B(T)d\lambda = \int\qa\J d\lambda\,,
}
where $q\sub{a\lambda} = \q(1 - \al)/(1 - \al_0)$ is the absorption efficiency
normalized to the fiducial wavelength $\lambda_0$. Because the scattering term
preserves scale invariance, it does not enter into this scale-setting equation
(which therefore becomes meaningless in the case of pure scattering, \al\ = 1).

The actual energy input into the dust is external radiation from a nearby
source, such as a star, galactic nucleus, etc., of radius \re\ and intensity
\Ie. When this source is non-spherical (for example disk or torus), the angular
profile of its geometrical shape, too, must be specified. The dust distribution
has some prescribed arbitrary geometrical shape and we denote the distance of
its closest point to the source of radiation $r_1$.  From the previous section,
the radiative transfer problem can not depend separately on \re\ or $r_1$, only
on their ratio
\eq{
                       \theta_{e1} = {\re \over r_1}.
}
Therefore, the only geometrical property of the radiation source relevant to
the problem is its angular size at the location of the dust (and its angular
shape profile when it is non-spherical). The radiative input to the dust can be
characterized by the external flux at $r_1$,
\eq{
   F\sub{e\lambda}(r_1) = \int\mu\Ie d\Omega \approx \upi\theta_{e1}^2\Ie.
}
This spectral distribution can be specified instead by its scale, the
bolometric flux at $r_1$
\eq{Fone}{
   \Fone = \int F\sub{e\lambda}(r_1) d\lambda = {\Le\over 4\upi r_1^2}\,,
}
where \Le\ is the luminosity, and spectral shape
\eq{
                      \fe = {F\sub{e\lambda} \over F_e}\,,
}
an $r$-independent normalized distribution ($\int\!\!\fe d\lambda = 1$).  Then
$F\sub{e\lambda}(r_1) = \fe\Fone$. Consider now the series of models in which
all properties are held fixed except the input bolometric flux \Fone.  Each
value of \Fone\ uniquely determines a corresponding value of $T_1$, the dust
temperature at $r_1$.  This in turn defines another flux $\sigma T_1^4$,
setting the intrinsic scale of dust emission. But the radiative transfer
problem can involve only one radiative scale, therefore the ratio of these two
fluxes must be a characteristic dimensionless function of the model. That is,
the ratio 
\eq{Fscaling}{
                     \Psi = {4\sigma T_1^4 \over \Fone}
}
is uniquely determined by the dimensionless properties of the model such as 
optical depths, angular shapes and dimensionless density profile (the factor 4 
is introduced for convenience). It may also depend on ratios of opacity spectral 
averages involving the profile \fe\ of the input radiation and
\eq{
                     \b(T_1) = {\upi\over\sigma T_1^4}\,\B(T_1),
}
the normalized ($\int\b d\lambda = 1$) spectral shape of the Planck
distribution at temperature $T_1$.  However, it can not depend on any
dimensional quantity such as luminosity, linear size or density.  In Section \ref{implications} we present results for $\Psi$ in spherical symmetry (figure 1) and derive an analytic approximation for its $\tau$-dependence (equation \ref{fit}).

This discussion shows that the luminosity never enters into the formulation of 
the radiative transfer problem, only the input flux matters.  And rather than 
specifying the flux scale, the dust temperature can be specified instead so that 
flux becomes a derived quantity.  {\em A dusty region with prescribed properties 
can transport only a single, well defined radiative flux for a given dust 
temperature on its heated boundary}. It is determined from the full solution 
through the scaling function $\Psi$, which can be viewed as an eigenvalue of 
the model.  The fact that the radiative transfer problem can be fully specified 
in terms of temperatures rather than radiative flux was first utilized in RR for 
spherical shells. Equation \ref{Fscaling} is the general form of this property 
for all geometries.

Dust sublimation occurs at some temperature \Tsub, determined by the grain
properties.  As long as the input flux \Fone\ is sufficiently low that $T_1 <
\Tsub$, the dust temperature at the inner boundary varies according to
eq.~\ref{Fscaling} ($T_1 \propto \Fone^{1/4}$) and different luminosities
produce dust emission that differs both in overall luminosity and spectral
shape.  This is generally the situation in planetary nebulae.  However, once
\Fone\ increases to the point that $T_1 = \Tsub$, the dust temperature cannot
rise any more. Instead, further increases in luminosity move out the inner
boundary of the dust so that \Fone\ remains fixed at the value that produces
$T_1 = \Tsub$. Thus the temperature profile and emission of the dust are set
by its internal properties in this regime. Stated differently -- since \Fone\
uniquely determines $T_1$, the converse is also true and $T_1$ uniquely
determines a corresponding \Fone. Models in which the dust is as close to the
radiation source as possible have the same $T_1$ and therefore adjust their
inner boundary to accommodate the external luminosity so that they also have
the same \Fone.  Since the bolometric flux is now self-regulated by dust
sublimation, {\em when the dust is as close to the radiation source as
possible, the only relevant property of the input radiation is its spectral
shape \fe}.

In this discussion we assumed single-component dust and heating by a single
external radiative source.  Multi-component dust mixtures are discussed in
Appendix A which shows that, although the technical complications increase,
all the essential scaling properties are preserved.  The same applies to
heating by more than one source.  Because radiative transfer can involve only
one intensity scale, the addition of an external source can only introduce the
ratio of corresponding fluxes $F\sub{e\lambda}$ at every point.

\section                      {SPHERICAL SYMMETRY}
\label{spherical}

To further explore concrete consequences of general scaling we must specify the 
geometry.  Here we take the case of spherical symmetry.  The essence of most 
results is preserved in other geometries.

In the case of spherical symmetry, $y$ can be taken as dimensionless distance 
from the center of symmetry, and \tT\ and $\eta(y)$ defined along the radial 
path. The radiation source is surrounded by a spherical dust shell with inner 
radius $r_1$ corresponding to dust sublimation, which we choose as the scaling 
length in equation \ref{rscaling}.  With this choice, dust sublimation always 
occurs at $y = 1$ and the dust temperature scale is set by $T(y{=}1) \equiv T_1 
= \Tsub$. The dust optical properties are fully specified by the wavelength 
variation of the absorption efficiency \qa\ and the albedo $\al$. Its spatial 
distribution is described by
\eq{
                  \eta(y) = {\k(y) \over \int_1^\infty\k(y)dy},
}
the dimensionless variation of extinction coefficients in the radial direction.  
Note again that $\eta(y)$ does not introduce a new scale since it is normalized 
via $\int_1^\infty\eta(y)dy = 1$.  When the dust properties do not vary with 
distance beyond the sublimation point, $\eta$ describes the radial law of 
density variation, e.g., $y^{-2}$, $\exp(-y)$, etc. There is no need to specify 
the actual density itself.  The extinction scale is set by $\tau_0$, the total 
optical depth along radial rays at the fiducial wavelength $\lambda_0$.  At all 
other wavelengths, $\tT = \tau_0\q$. At any point $y$, the optical depth from 
the closest approach to the center along a path that makes an angle $\theta$ 
with the radius vector is
\eq{
                  \t(y,\theta) = \tT\int_0^{y\cos\theta}\!\!\!\!\!
                  \eta\left(\sqrt{u^2 + y^2\sin^2\theta}\right) du.
}
Rowan-Robinson was the first to incorporate geometrical scaling for spherical
shells into a formalism that specifies only scaled radii, \tT\ and $\eta$.

From the formal solution of the radiative transfer equation, the intensity at
radius $y$ of a ray inclined at angle $\theta$ to the radial direction is
\eq{formal}{
  \I(y,\theta) =
   \Ie e^{-\t(y)}\Theta\!\left({\theta_{e1}\over y} - \theta\right)
   + I\sub{d\lambda}(y,\theta)}.
Here $\Theta$ is the step function (unity for positive arguments, zero
otherwise) and
\eq{
I\sub{d\lambda}(y,\theta) =
 \int\!\! \S(y',\theta) e^{\t(y',\theta) - \t(y,\theta)}d\t(y',\theta)
}
is the intensity of the diffuse radiation. By angular integration, the
angle-averaged intensity is similarly
\eq{J}{
          \J(y) = {\Fone\over 4\upi y^2}\fe e^{-\t(y)} + \Jd(y),
}
where \Jd\ denotes the diffuse component. We now show that \Jd, which is
comprised of dust emission and scattering, can be expressed explicitly in terms
of \J. For the emission term this result is straightforward since this term is
controlled by the dust temperature $T$, determined from
\eq{temp}{
\qaP(T)T^4 = \qaP(T_1)T_1^4 \,{\int\qa\J(y)d\lambda\over\int\qa\J(1)d\lambda}
}
(cf equation \ref{equilib}). Here we introduced the Planck mean of the
absorption efficiency
\eq{
                    \qaP(T) = \int\qa\b(T)d\lambda.
}
For the scattering component, consider first the case of isotropic scattering,
$g(\Omega',\Omega) = 1$. Then the scattering term in the source function is
simply \al\J\ (eq.~\ref{source}), and the entire diffuse component \Jd\ becomes
an explicit function of \J.  Similarly, for the external radiation, radiative
equilibrium at $y = 1$ (where $T = T_1$) determines the input bolometric flux
\Fone, or equivalently $\Psi$ (see eq.~\ref{Fscaling}), as
\eq{Psi}{
         \Psi = {\qae\over\qaP(T_1)}{1\over 1 - \epsilon}
}
where
\eq{qe}{
          \qae = \int\qa\fe d\lambda\,, \qquad
          \epsilon = {\int\qa\Jd(1)d\lambda \over \int\qa\B(T_1)d\lambda}\,.
}
Since \Jd\ is an explicit function of \J, the same holds for $\Psi$. Therefore,
the entire right-hand-side of equation \ref{J} is expressed explicitly in terms
of \J, so this is a self-consistency equation for the unknown \J\ whose only
inputs are dust properties and the spectral shape of incoming radiation. As an
inhomogeneous integral equation, it has a unique physical solution. This
completes the formulation of the radiative transfer problem in accordance with
the general scaling discussion of the previous section. The only property of
the external radiation that needs to be specified as input is its spectral
shape \fe.

Anisotropic scattering can be handled similarly. In that case we write
$\I(\theta) = \Phi(\theta)\J$, where $\Phi$ is the intensity angular
distribution, normalized as $\int\Phi d\Omega = 4\upi$.  Equation \ref{J} for
\J\ remains the same, only the scattering contribution to \Jd\ includes now an
integral over $\Phi$.  The only technical complication is the need to
additionally solve the coupled equation (\ref{formal}) for the angular
distribution $\Phi$. The rest of the formulation is the same as for isotropic
scattering.  In complete analogy, arbitrary geometries can be handled through
equally straightforward extensions.

Solution of equation \ref{J} for \J\ determines the radial variation of the
source function in the shell.  Once this function is known, the flux and
surface brightness, the quantities relevant for observations, are readily
calculated. In particular, the overall flux is obtained from the first-moment
angular integration of equation \ref{formal} which yields
\eq{F}{
            \F(y) = {\Fone\over y^2}\fe e^{-\t(y)} + \Fd(y),
}
similar to the expression for \J.  As with the external radiation, it is
convenient to remove the overall scale, contained in the bolometric flux
$F = \int\F d\lambda$.  The spectral shape of the overall flux is denoted \f,
so that
\eq{
          \F(y) = \f(y)F(y), \qquad    F(y) = {F_1\over y^2}.
}
Here $F_1$ is the bolometric flux at $y = 1$, including both external and dust
contributions, and the second equality expresses flux conservation.

The luminosity \Le\ has never entered, and is irrelevant.  It is important to 
note, however, that the bolometric flux $F_1$ is fully determined even though 
the overall luminosity is not.  This flux combines the dimensional scales \Le\ 
and $r_1$ which otherwise do not enter individually, thus dropping out of the 
problem. The corresponding relation, obtained by combining equations \ref{Fone} 
and \ref{Fscaling}, can be written as
\eq{r1}{
            r_1^2 = {\Le\over16\upi\sigma T_1^4}\Psi
                  = {\Psi\over4}\rre\left({\Teff\over T_1}\right)^{\!4},
}
where the last equality is written in terms of the source's effective
temperature \Teff.  This is the exact expression of a result noted in various
forms by RR, Laor \& Draine (1993) and Ivezi\' c \& Elitzur (1996a; IE96
hereafter). Therefore, when the overall luminosity is known,  it can be used to
determine the radius of the dust sublimation zone.

The only properties of the central source that affect the observed radiation
are its spectral shape \fe\ and angular size $\theta_{e1}$.  The latter
enters explicitly only in the expression for the intensity, eq.~\ref{formal},
thus it can be determined directly only in high-resolution observations that
delineate the external contribution as a bright spot.  In most cases, though,
the central source is unresolved and the quantity relevant for observations is
the flux.  The angular integration that transforms intensity to flux removes
the explicit dependence on $\theta_{e1}$, as is evident from eq.~\ref{F}, but
the flux may still depend on $\theta_{e1}$ because of occultation of the
diffuse radiation by the central source.  Without occultation, the diffuse flux
would vanish at $r_1$ because the intensities of the two streams along any ray
are then the same (Milne, 1921).  Thus the significance of occultation can be
gauged from the ratio $F_d(r_1)/\Fone$ of the diffuse and external
contributions to the bolometric flux at $r_1$.  Only rays occulted by the
central source contribute to $\Fd(r_1)$, and $\B(T_1)$ is an upper limit to
their intensity.  Therefore, $\Fd(r_1) < \upi\theta_{e1}^2\B(T_1)$ at all
wavelengths and
\eq{
      {F\sub d(r_1)\over\Fone} < \left({T_1\over\Teff}\right)^{\!4}.
}
Occultation can be ignored whenever $T_1 \ll \Teff$. Comparison with
eq.~\ref{r1} shows that this condition is equivalent to $\theta_{e1}^2\Psi
\ll 1$. When this condition is obeyed,
\eq{Ffin}{
      F_1 = \Fone,  \qquad \f(y) = \fe e^{-\t(y)} + y^2\Fd(y)/F_1,
}
and the problem is independent of $\theta_{e1}$. Therefore, as long as $\Teff 
\gg \Tsub = T_1$, the only relevant property of the central source is its 
spectral shape \fe; \Teff\ is irrelevant, and in the remainder of the discussion 
we assume that this condition is fulfilled.\footnote{For a more detailed 
discussion of backwarming effects see Rowan-Robinson (1982).}  When \Teff\ 
decreases and approaches \Tsub, a dependence on angular size $\theta_{e1}$ 
enters, but \Teff\ itself remains irrelevant as long as $T_1 = \Tsub$ is 
maintained.  Finally, scaling breaks down when \Teff\ is so small that $T_1 < 
\Tsub$.  In this case, \Teff\ must be specified, too, since it determines $T_1$. 
This breakdown of scaling occurs also when the central cavity is dust-free 
because of factors other then dust sublimation, such as dynamics for example, 
and $T_1 < \Tsub$.

\section                      {IMPLICATIONS}
\label{implications}

When the dust is optically thin at all wavelengths, the solution is
straightforward.  Inserting $\tau_\lambda \sim 0$ and neglecting the diffuse
emission, equation \ref{J} gives $\J(y) = \J(1)/y^2$ and so
\eq{thin}{
  {\qaP(T)\over\qaP(T_1)}\left(T \over T_1\right)^{\!\!4} = {1\over y^2},
  \qquad
  \Psi = {\qae \over \qaP(T_1)}\,.
}
Most of the contribution to the Planck mean comes from the distribution peak
around
\eq{lambdaT}{
                  \lp(T) = 4\,\mic\,{1000\, \rm K \over T},
}
so $\qaP(T) \simeq q\sub{a\lambda_p(T)}$. If the corresponding wavelength is in
the regime where the absorption efficiency drops as a power law, $\qa \propto
\lambda^{-\beta}$, we recover the familiar result for the dust temperature $T
=T_1 y^{-2/(4 + \beta)}$.  Also, when the spectral shape of the external
radiation is that of a black-body with temperature \Te, the optically thin
limit of $\Psi$ becomes
\eq{Psi0}{
 \Psi(\tau = 0) \simeq {q\sub{a\lambda_p(T_e)} \over q\sub{a\lambda_p(T_1)}};
 }
that is, it is determined by the drop in absorption efficiency between the
wavelengths corresponding to the temperatures of the external source and dust
sublimation.

When the dust is optically thick the problem must be solved numerically, and we 
have developed a computer code (DUSTY; Ivezi\'c, Nenkova \& Elitzur 1997) to 
solve the spherical problem.  A description of the numerical procedure is 
provided in Appendix C.  Our program solves equation \ref{J} for $\J(y)$, with 
the dust temperature determined from equation \ref{temp} and the flux scale $F_1 
= 4\sigma T_1^4 \Psi$ from equation \ref{Psi}.  Input includes the optical 
properties \qa\ and $\al$, the dust sublimation temperature \Tsub, its density 
distribution $\eta$ and the external spectral shape \fe.  A given set of these 
quantities defines a family of solutions, distinguished from each other by 
overall optical depth. Matching a model to observations provides the optical 
depth of the source if there are independent estimates for the other properties. 
Note that the only effects on the spectral energy distribution that can be 
meaningfully discerned in observations involve changes of spectral features or 
of slope.

We present now the results of exact calculations, first for what we will use
as a ``standard" model with $\eta = y^{-2}$, single-component dust grains with
size $a$ = 0.05 \mic\ and sublimation temperature \Tsub\ = 700 K, and external
radiation with the spectral shape of a black body with \Te\ = 2500 K.  We then
explore the effects of different input quantities on the emergent flux and
discuss the behavior of the solutions with the aid of analytic approximations
developed in Appendix B.  In all cases we vary \tV, the overall optical depth
at 0.55 \mic, all the way to 1000 and present solutions for both amorphous
carbon and silicate grains.  Optical properties for the former are taken from
Hanner (1988), for the latter from Ossenkopf, Henning \& Mathis (1992).  The
scattering is assumed isotropic.  Similar explorations of parameter space were
performed by Leung and by Rowan-Robinson.  In addition to employing updated
optical properties, the coverage presented here is more extensive and the
systematics of the presentation more closely reflect scaling.

%%%%%%%%%%%%%%%%%%% Figure 1 %%%%%%%%%%%%%%%%%%%%%%%%%%%%%
\begin{figure}
\centering \leavevmode \figsize \epsfclipon \epsfbox[70 160 505 580]{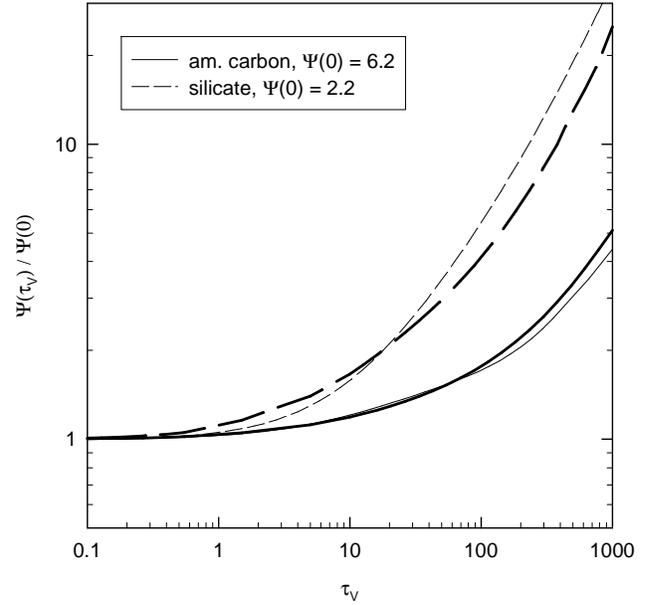}

\caption{Variation of the scaling function $\Psi$ (equation 15) with overall
visual optical depth \tV\ in the ``standard model" (see text).  Thick lines
are the results of exact numerical calculations, the thin lines are the
analytical approximations of equation B7.}

\end{figure}
%%%%%%%%%%%%%%%%%%%%%%%%%%%%%%%%%%%%%%%%%%%%%%%%%%%%%%%%%%%%%%%%%%%%%%%%%%%%

Figure 1 displays the variation of $\Psi$ with \tV.  For each chemical
composition, the optically thin value is a measure of the drop in absorption
between $\lp(\Te) \simeq 2 \mic$ and $\lp(\Tsub) \simeq 6 \mic$ (equation
\ref{Psi0}).  The thin lines are the analytic results of equation
\ref{approx}, seen to provide excellent approximations to the exact
calculations.  These analytic expressions are not too transparent as they
involve spectral integrations of the detailed extinction efficiencies.
Therefore, we discuss the results instead in terms of the gray-opacity
approximation, equation \ref{gray}, which maintains most of the main features.
In particular, it explains the rise of $\Psi$ with \tV\ at large optical
depths.\footnote{Formally, this shows that $\Psi \to \infty$ when $\tV \to
\infty$ with all other properties held fixed, and the inner radius of the
shell increases without bound (equation \ref{r1}). However, the solutions are
derived assuming steady-state and the time to reach this stage diverges too in
that limit.} Indeed, the numerical results are well described by the simple
analytical fit
\eq{fit}{
         {\Psi(\tV) \over \Psi(0)} \simeq 1 + 0.005\,\tau_{\rm V}^m ,
}
where the power $m$ is 1 for amorphous carbon and 1.25 for silicate grains.
This provides an adequate approximation for $\Psi$ of the ``standard model"
over the entire displayed range, the accuracy is better than 20\% for
amorphous carbon, 40\% for silicates.

%%%%%%%%%%%%%%%%%%% Figure 2 %%%%%%%%%%%%%%%%%%%%%%%%%%%%%
\begin{figure}
\centering \leavevmode \figsize \epsfclipon \epsfbox[70 100 500 720]{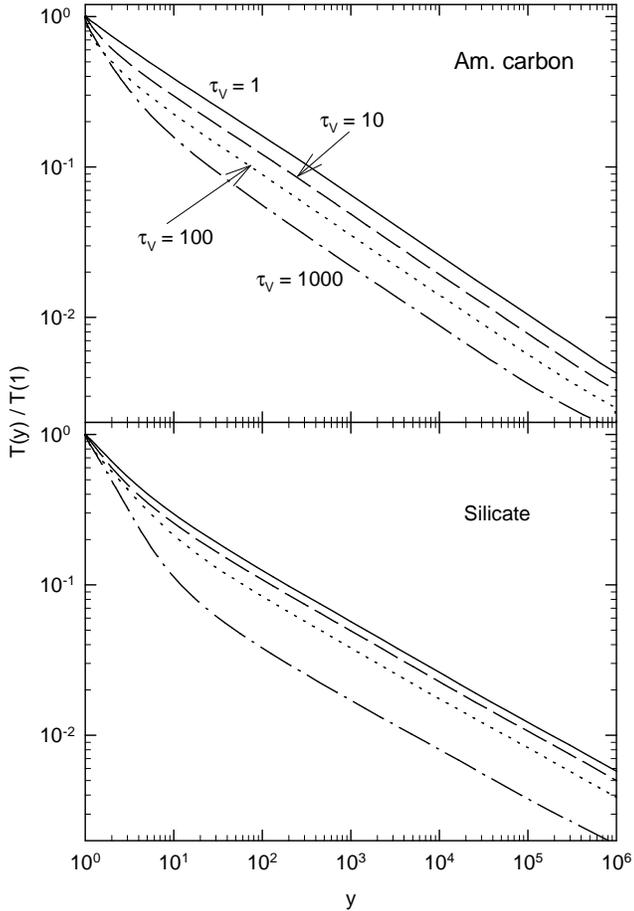}

\caption{The radial variation of dust temperature in the ``standard model" for
various optical thicknesses, as indicated.}

\end{figure}
%%%%%%%%%%%%%%%%%%%%%%%%%%%%%%%%%%%%%%%%%%%%%%%%%%%%%%%%%%%%%%%%%%%%%%%%%%%%

Figure 2 shows the spatial temperature profile for various \tV.
With the density profile $\eta = y^{-2}$, the gray-opacity result gives a
temperature drop $T \propto y^{-3/4}$ in the inner regions $y \la \tV$,
followed by a more moderate decline $T \propto y^{-1/2}$ at larger radii.
Qualitatively, this is the behavior displayed in the figure, the actual
absorption efficiencies only modify the details.

%%%%%%%%%%%%%%%%%%% Figure 3 %%%%%%%%%%%%%%%%%%%%%%%%%%%%%
\begin{figure}
\centering \leavevmode \figsize \epsfclipon \epsfbox[65 100 500 720]{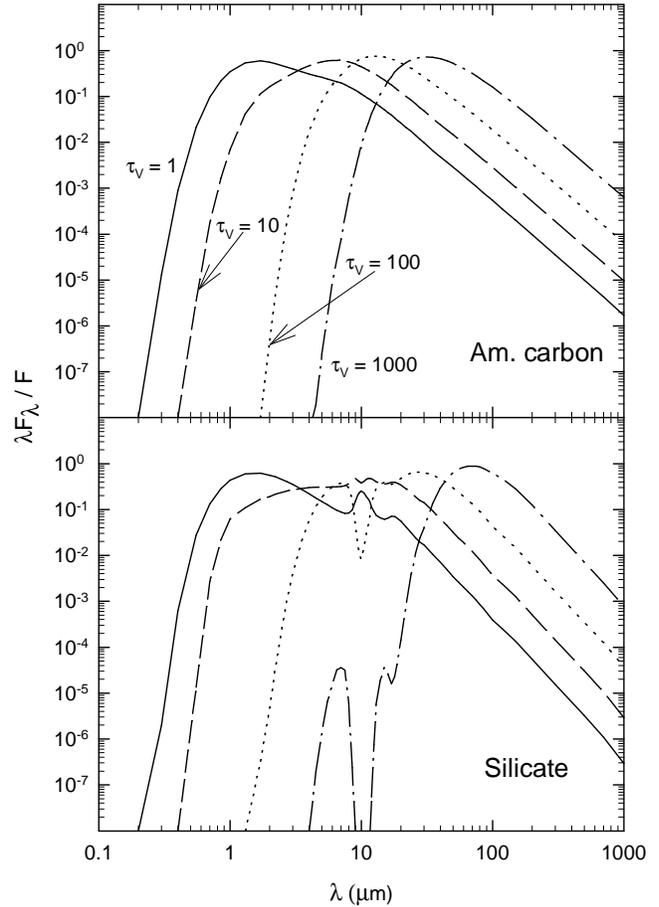}

\caption{The spectral energy distribution of the ``standard model" for various
optical thicknesses, as indicated.}

\end{figure}
%%%%%%%%%%%%%%%%%%%%%%%%%%%%%%%%%%%%%%%%%%%%%%%%%%%%%%%%%%%%%%%%%%%%%%%%%%%%

Figure 3 displays the variation of the SED with overall optical depth. The
shape of the SED is governed by some general characteristics, common to all
solutions. Because the dust cannot be warmer than \Tsub, dust emission is
negligible for $\lambda \la \lp(\Tsub)$ and scattered light dominates the
diffuse component at these wavelengths.  As \tV\ increases beyond \about\ 10,
the external radiation and scattered light are completely absorbed and dust
emission dominates at all relevant wavelengths.  At each wavelength $\lambda$,
dust radiation is generated at radius $y$ where $T(y) \sim
(40\,\mic/\lambda)\x100$ K (equation \ref{lambdaT}). For optically thick
configurations, Kwan \& Scoville (1976) noted that the SED peaks at the
wavelength corresponding to \t\ \about\ 1.  At longer wavelengths, the source
is always optically thin and the spectral shape is independent of overall
optical depth.  In this region, the shape of the SED reflects only the
long-wavelength behavior of the absorption coefficient and the density
distribution.  The net result of all these effects is that in optically thick
shells, the SED roughly retains its shape, shifting as a whole to longer
wavelengths as \tV\ increases.

This qualitative discussion adequately explains the general behavior of
spectral shapes. We proceed now to explore the effects of different input
properties on the SED, varying them one by one over the entire plausible
range; the properties not varied are those of the ``standard" model.

%%%%%%%%%%%%%%%%%%% Figure 4 %%%%%%%%%%%%%%%%%%%%%%%%%%%%%
\begin{figure}
% \begin{minipage}{1.8\hsize}
\centering \leavevmode \figsize \epsfclipon \epsfbox[55 40 585 766]{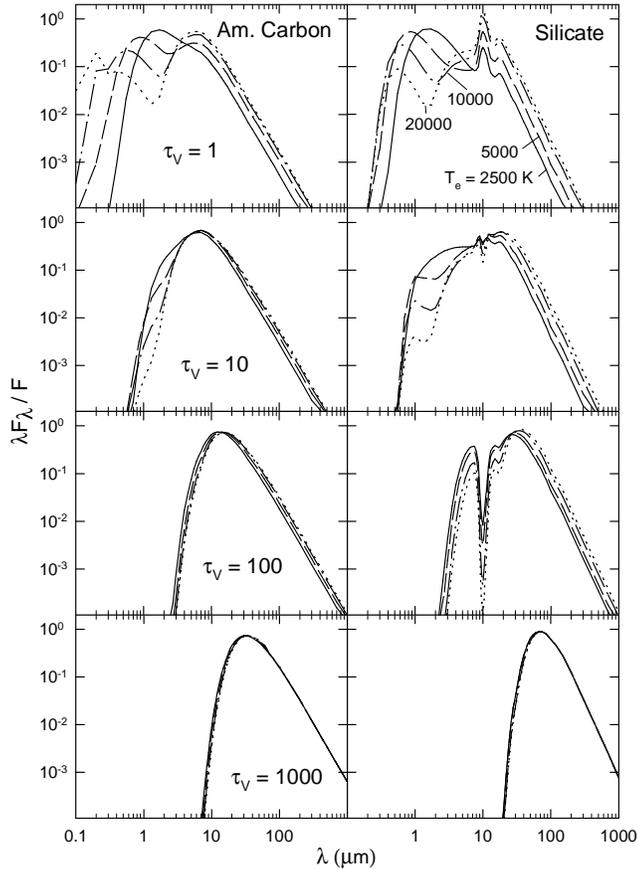}

\caption{Effect of the external radiation spectral shape on the emerging SED
for various optical thicknesses, as marked.  The external spectral shape is
that of a black-body with temperature \Te\ as indicated in the top right
panel.  In this and all subsequent figures, the fixed properties are those of
the ``standard model".}

% \end{minipage}
\end{figure}
%%%%%%%%%%%%%%%%%%%%%%%%%%%%%%%%%%%%%%%%%%%%%%%%%%%%%%%%%%%%%%%%%%%%%%%%%%%%

\subsection {External spectral shape \fe}

When the external radiation has a black-body spectral shape with a temperature 
\Te, $\fe = \b(\Te)$. figure 4 displays the effect of varying \Te\ from 2500 K 
to 20000 K at various optical thicknesses. In all cases, longward of 10 \mic\  
\Te\ has virtually no effect on the observed spectral shape.  The only minor 
effect of \Te, with no practical observational implications, is to slightly 
modify the relative scale of emission.  The weak dependence on \Te\ was also a 
result of RR. The reason \fe\ has no effect at long wavelengths is that the 
external contribution is negligible there in comparison with the dust emission; 
although the source is brighter, the dust emission covers a much larger surface 
area.  The external radiation emerges at short wavelengths because dust emission 
disappears at $\lambda \la \lp(\Tsub)$. However, even in that spectral region 
the external radiation is visible only when $\tV \la 1$. At larger optical 
depths it is fully absorbed, becoming irrelevant at all wavelengths.  Therefore, 
the external spectral shape can be determined only in short-wavelength 
observations when the optical thickness is not too large. This conclusion 
applies to all external spectral shapes, not just those that follow Planck's 
law.

%%%%%%%%%%%%%%%%%%% Figure 5 %%%%%%%%%%%%%%%%%%%%%%%%%%%%%
\begin{figure}
\centering \leavevmode \figsize \epsfclipon \epsfbox[55 40 585 766]{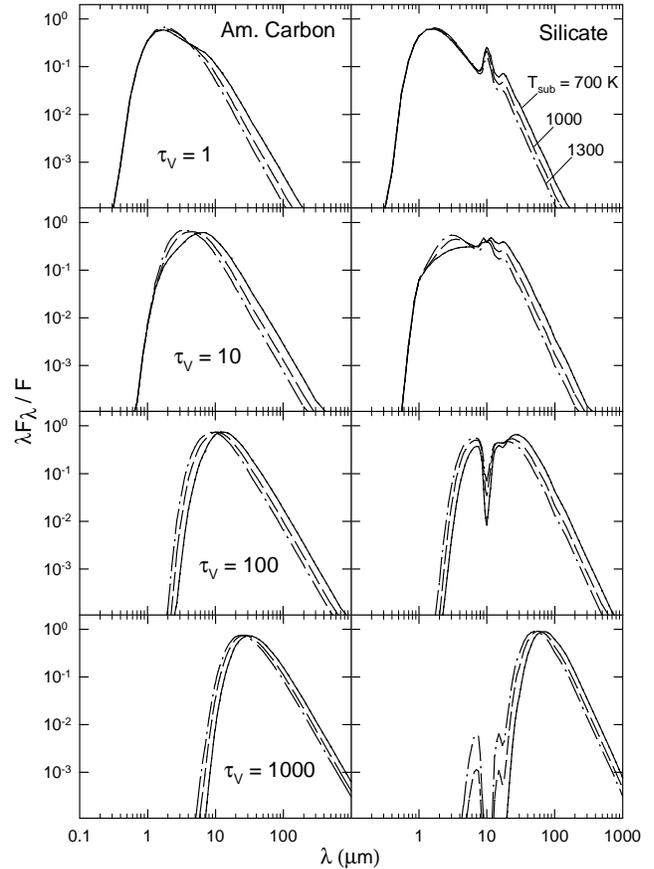}

\caption{Effect of dust sublimation temperature \Tsub, indicated in the top
right panel, on the emerging SED.}

\end{figure}
%%%%%%%%%%%%%%%%%%%%%%%%%%%%%%%%%%%%%%%%%%%%%%%%%%%%%%%%%%%%%%%%%%%%%%%%%%%%

\subsection {Dust sublimation temperature \Tsub}

Figure 5 displays the effect of \Tsub\ on the observed spectral shape. As
expected, \Tsub\ variations have no effect on the shape of the SED at
wavelengths shorter than \lp(\Tsub).  At longer wavelengths, varying \Tsub\
affects only the scale of the SED not its shape, similar to the effect caused
at those wavelengths by variations of \Te.  Furthermore, at large optical
depths ($\tV \ga 100$), spectra produced by varying \Tsub\ are
indistinguishable at all wavelengths from those produced by varying \tV.  As a
result, uncertainty in \Tsub\ translates into an uncertainty in the optical
depth obtained from a fit to the observed SED.

Because the impact of \Tsub\ on the spectral shape is minor and poorly
separated from other effects, \Tsub\ cannot be well determined from the SED.
The only regime where such determination can be realistically attempted is
intermediate optical depths, \tV\ \about\ 10. In general, the dust sublimation
temperature is more reliably determined in high-resolution observations
whenever they are capable of resolving the dust condensation zone. From
equation \ref{Fscaling},
\eq{
         \sigma T_1^4 = \Psi {F\sub{obs} \over \theta^2_{\rm 1,obs}}
}
where $F\sub{obs}$ and $\theta_{\rm 1,obs}$ are, respectively, the observed
flux and dust condensation zone angular diameter (see also IE96). Whenever
feasible, this is the more practical method and we have successfully employed
it for analysis of IRC+10216 (Ivezi\' c \& Elitzur, 1996b).

%%%%%%%%%%%%%%%%%%% Figure 6 %%%%%%%%%%%%%%%%%%%%%%%%%%%%%
\begin{figure}
\centering \leavevmode \figsize \epsfclipon \epsfbox[55 40 585 766]{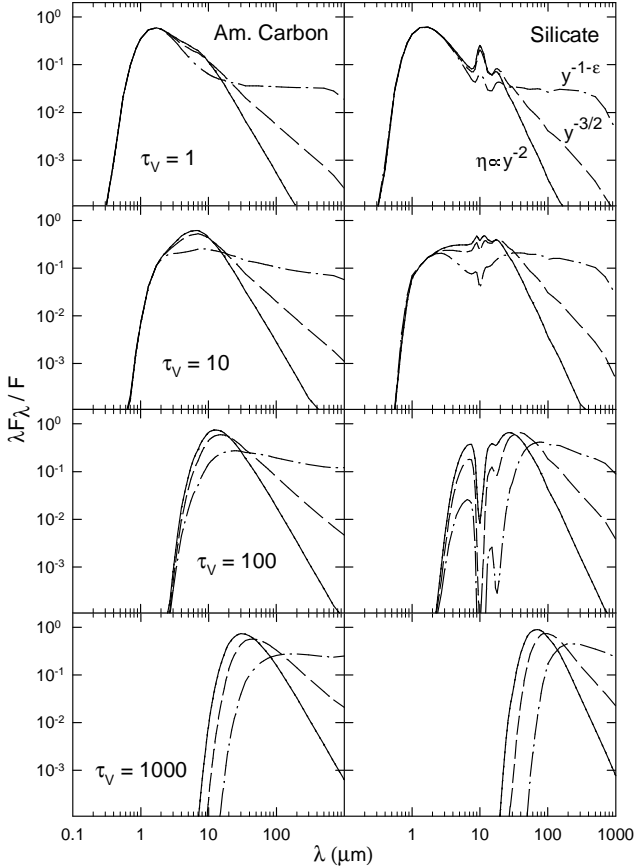}

\caption{Effect of dust density profile $\eta$ on the emerging SED for
power-law profiles steeper than $y^{-1}$, indicated in the top right panel.}

\end{figure}
%%%%%%%%%%%%%%%%%%%%%%%%%%%%%%%%%%%%%%%%%%%%%%%%%%%%%%%%%%%%%%%%%%%%%%%%%%%%

\subsection {Dust distribution $\eta$}

Figures 6 -- 8 display the effect of the dust density distribution, presenting
the results for power-law variation $\eta \propto y^{-p}$ for various
non-negative $p$.  As is evident from these figures, for given dust optical
properties $\eta$ is the input quantity with the most substantial impact on
the long wavelength emission, the only one to produce an unambiguous
observational signature.  Different density distributions can be easily
distinguished by colors involving $\lambda \ga$ 10 \mic.

Varying the dust density distribution affects the emerging spectrum in two
ways.  It changes the amount of dust responsible for the emission at a given
wavelength and it also modifies the temperature profile.  However, the
differences between the temperature profiles play only a secondary role to
those in the matter distribution itself.  The flux at wavelength $\lambda$ is
dominated by emission from radius $y$ where the dust temperature $T$ obeys
$\lp(T) \sim \lambda$, so redistributing the material between the various
radii directly modifies the spectral shape.  Power-law distributions with $p
\le 1$ must include a cutoff to ensure finite column density.  We therefore
make a distinction between two families of distributions, discussed separately.

\subsubsection {Steep distributions; $p > 1$}

Steep density distributions do not require a cutoff. Figure 6 displays the
SEDs for $p$ = 2, 3/2 and $1 + \epsilon$, where $\epsilon \ll 1$.  The
spectral shapes become flatter at long wavelengths as $p$ decreases since for
the same overall optical depth, relatively more material is placed at larger
radii (i.e., low temperatures).  In particular, for power-law variations $\qa
\propto \lambda^{-\beta}$ and $\eta \propto y^{-p}$, Harvey et al.\ (1991)
show that $\lambda\f \propto \lambda^{-(\beta + 4)(p - 1)/2}$ in the optically
thin regime.  We find this relation useful for qualitative understanding of
the SED past its peak.  Since the differences among SEDs reflect dust
emission, these differences disappear at $\lambda \la \lambda(\Tsub)$. Indeed,
as the figure shows, the spectral shapes are indistinguishable at such
wavelengths, therefore the density distribution cannot be determined from the
observed SED at $\lambda \la \lambda(\Tsub)$. At large optical depth, $\tV \ga
50$, the emission peak shifts to wavelengths longer than $\lambda(\Tsub)$ and
the SEDs differ from each other on both sides of the peak.

%%%%%%%%%%%%%%%%%%% Figure 7 %%%%%%%%%%%%%%%%%%%%%%%%%%%%%
\begin{figure}
\centering \leavevmode \figsize \epsfclipon \epsfbox[55 40 585 766]{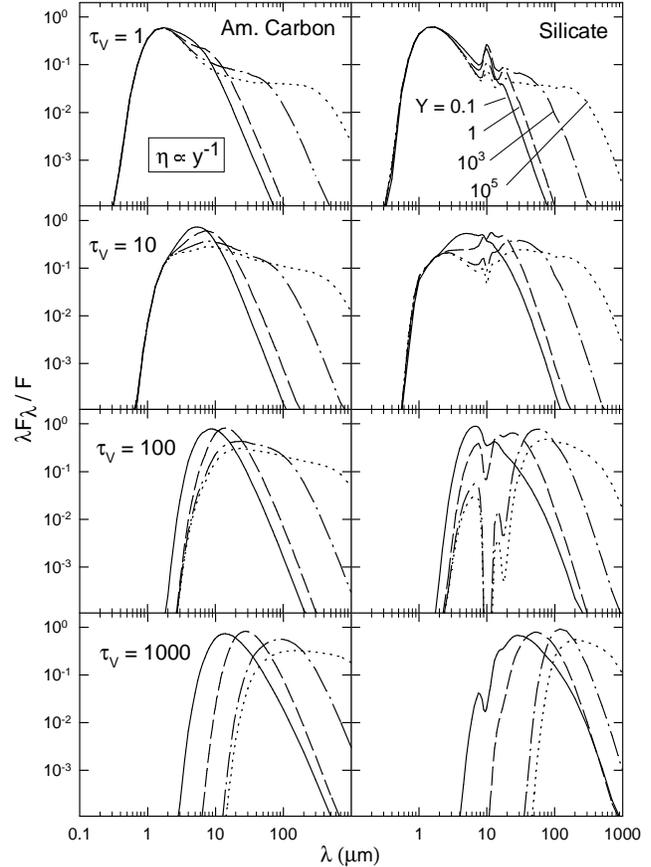}

\caption{Effect of the shell's geometrical thickness on the emerging SED for
dust density profile $\eta \propto y^{-1}$.  Only the relative thickness $Y =
\Delta r/r_1$, indicated in the top right panel, is relevant}

\end{figure}
%%%%%%%%%%%%%%%%%%%%%%%%%%%%%%%%%%%%%%%%%%%%%%%%%%%%%%%%%%%%%%%%%%%%%%%%%%%%

%%%%%%%%%%%%%%%%%%% Figure 8 %%%%%%%%%%%%%%%%%%%%%%%%%%%%%
\begin{figure}
\centering \leavevmode \figsize \epsfclipon \epsfbox[55 40 585 766]{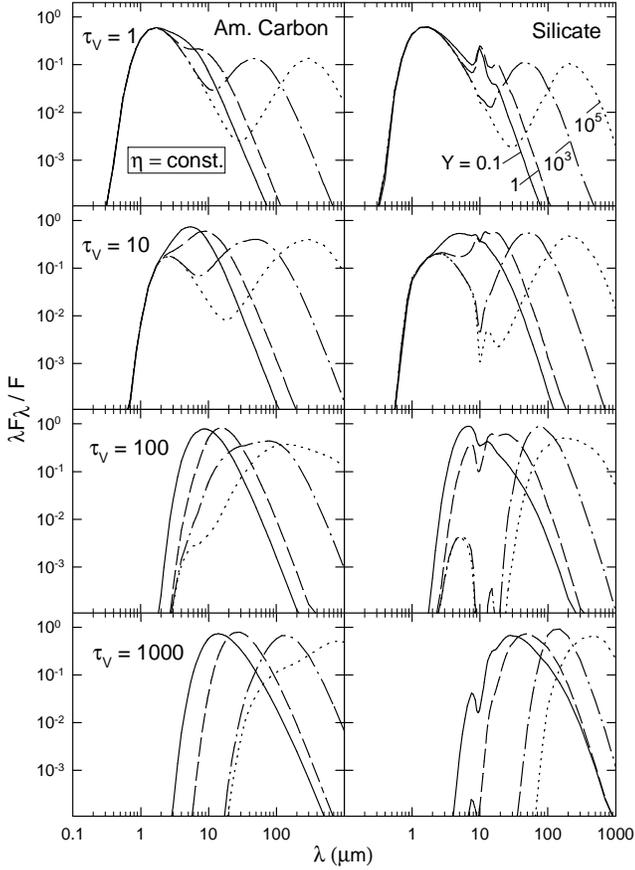}

\caption{Same as figure 7, but for a constant density distribution.}

\end{figure}
%%%%%%%%%%%%%%%%%%%%%%%%%%%%%%%%%%%%%%%%%%%%%%%%%%%%%%%%%%%%%%%%%%%%%%%%%%%%

\subsubsection {Flat distributions; $p \le 1$}

Power-law distributions with $p \le 1$ require a cutoff \yout\ to ensure a
finite column density, introducing an additional independent parameter, the
shell relative thickness $Y = \yout - 1$.   This introduces in turn a lower
limit for the temperature and a corresponding wavelength $\lout =
\lp(T(\yout))$.  Dust emission is mostly confined to wavelengths shorter that
\lout, which thus serves as an effective long-wavelength cutoff for the SED.

The cutoff effect is best seen by comparing the SEDs for $\eta \propto y^{-1}$
of figure 7 to those displayed in figure 6 for $\eta \propto y^{-(1 +
\epsilon)}$. The latter density distribution does not require a $y$-cutoff and
displays the intrinsic effect of this power law for an infinite shell.  The
strict $y^{-1}$ distribution must include a cutoff, which is varied in figure
7 over a wide range.  For all practical purposes, the SED for $Y = 10^5$ is
indistinguishable from that for the infinite shell. The difference between the
two is noticeable only at wavelengths longer than \about\ 300 \mic, where the
SED of the finite shell dips down.  As the cutoff $Y$ decreases, \lout\
decreases too as more cold dust is removed from the shell and the decline of
the SED is pushed to shorter wavelengths.

Figure 8 presents the SEDs for constant-density shells with various
thicknesses. The gray-opacity result of equation \ref{gray} yields
\eq{
             \lout \propto \left[3\tV(Y + 1) + (Y + 1)^2\right]^{1/4}
}
for such shells when they are not too thin either physically ($Y \gg 1$) or
optically ($\tV \gg 1$).  This simple relation explains adequately the behavior
of the long-wavelength cutoff evident in the various panels of figure 8. When
$Y \ll \tV$, the cutoff wavelength varies in proportion to $(\tV Y)^{1/4}$.
When $Y \gg \tV$, the cutoff wavelength becomes independent of \tV\ and varies
in proportion to $Y^{1/2}$. In geometrically-thin ($Y \ll 1$) optically-thick
($\tV \gg 1$) shells, the cutoff wavelength varies in proportion to
$\tV^{1/4}$, independent of the shell geometrical thickness.

It is important to note that, although steep density distributions do not
require a formal cutoff, in practice such cutoffs always exist because of the
finite sizes of dust shells.  These finite sizes affect the long wavelength
part of the SED in exact analogy to the cutoff of a flat distribution.  The
finite size of a steep density distribution is relevant only for observations
at wavelengths longer than the corresponding \lout.  In all cases, the cutoff
effect can be easily masked by background emission because it generally
involves low temperatures.

%%%%%%%%%%%%%%%%%%% Figure 9 %%%%%%%%%%%%%%%%%%%%%%%%%%%%%
\begin{figure}
\centering \leavevmode \figsize \epsfclipon \epsfbox[65 290 600 572]{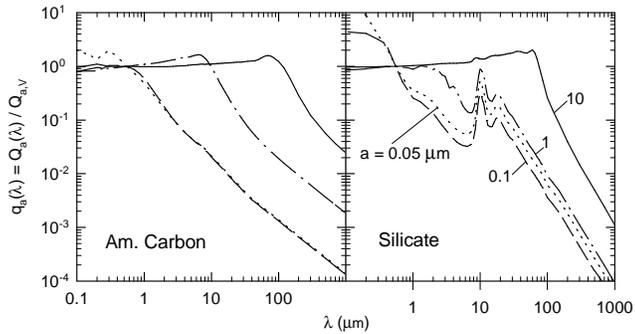}

\caption{Variation of the absorption efficiency spectral shape with grain
radius, indicated in the right panel.}

\end{figure}
%%%%%%%%%%%%%%%%%%%%%%%%%%%%%%%%%%%%%%%%%%%%%%%%%%%%%%%%%%%%%%%%%%%%%%%%%%%%

\subsection {Optical properties; grain chemistry and size}

The optical properties \qa\ and $\al$ depend on the grain chemical composition
and size.  The effect of composition on the SED can be discerned in all
figures by comparing the panels for amorphous carbon and silicate grains.  The
significance of grain size $a$ can first be gauged from figure 9.  For the two
chemical compositions and various radii $a$ it displays \qa, obtained from a
Mie theory calculation with the appropriate dielectric constants. Except for
spectral features, the extinction efficiency is a self-similar function of the
scaled variable $2\upi a/\lambda$, approximately constant at short wavelengths
and switching to a power-law decline with increasing $\lambda$ at long
wavelengths.  In this approximation, the shape of \qa\ is independent of grain
size for $a \la \lambda/2\pi$.  Therefore, at wavelengths longer than visual,
grain sizes are irrelevant as long as $a \la$ 0.05 \mic. On the other hand,
larger grain radii will significantly affect the results because they are
effectively equivalent to selecting grains with different radiative
properties.  Increasing the grain size beyond 0.05 \mic\ amounts to increasing
\qa\ at every wavelength longer than visual.  Therefore, for the same \tV,
increasing the grain size increases all far-IR optical depths.

%%%%%%%%%%%%%%%%%%% Figure 10 %%%%%%%%%%%%%%%%%%%%%%%%%%%%%
\begin{figure}
\centering \leavevmode \figsize \epsfclipon \epsfbox[55 40 585 766]{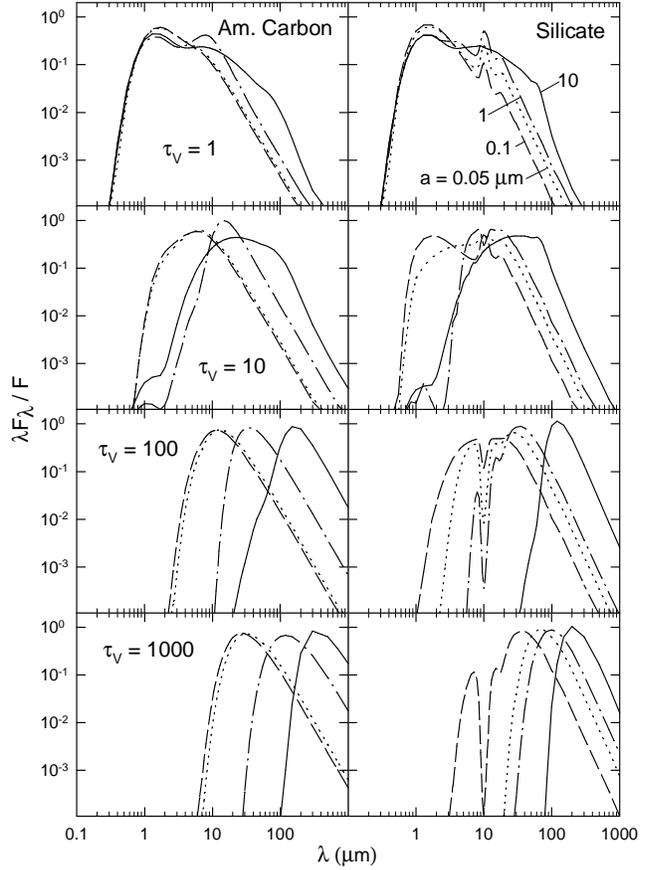}

\caption{Effect of grain size, indicated in the figure, on the emerging SED.}

\end{figure}
%%%%%%%%%%%%%%%%%%%%%%%%%%%%%%%%%%%%%%%%%%%%%%%%%%%%%%%%%%%%%%%%%%%%%%%%%%%%

Figure 10 displays the SEDs for our ``standard model" with grains of different
sizes and various optical depths.  The results for all grains smaller than
0.05 \mic\ should be the same as for $a$ = 0.05 \mic.  As could be expected,
the effect of grain size becomes significant only for $a >$ 0.1 \mic.  The
difference between the SEDs for $a$ = 0.05 and 0.1 \mic\ is generally small,
especially for amorphous carbon.  Larger grains, $a \ga$ 1\mic, generate some
structure at intermediate optical depths.  This structure, evident in the top
two panels for amorphous carbon, arises purely from optical depth effects,
reflecting the interplay between the stellar component and dust emission and
scattering.  It can give a false impression of dust spectral features,
although they do not exist in the actual absorption efficiency (see figure 9).

\section                         {DISCUSSION}

Scaling has far reaching consequences for modeling and analysis of IR emission 
from radiatively heated dust.  Thanks to scaling, the number of independent 
input properties is significantly reduced.  Only two scales need be specified 
for a complete solution --- the overall optical depth \tV\ and the dust 
sublimation temperature \Tsub.  All other input properties involve only 
dimensionless, normalized profiles.  For the external radiation, only the 
spectral profile \fe\ is required.  For the dust density, only the normalized 
distribution $\eta$ (equation \ref{eta gen}) is needed.  And for the dust 
optical properties, only the spectral shapes of absorption and scattering 
efficiencies enter.

Scaling enables a meaningful systematic study of the observational impact of
each independent input property on the emerging IR radiation. The results of
the systematic coverage of parameter space presented in the previous section
outline the range of spectral shapes produced by spherical configurations.
They show which properties delineated in observations can and cannot be
explained in this symmetry. For example, the IR spectral shape $\lambda\F
\propto \lambda^{-3/4}$ is sometimes used as an indicator of disk geometry
(e.g.\ Hillenbrand et al.\ 1992).  In fact, this spectral behavior is also
produced by spherical distributions with $\eta \propto y^{-p}$ with $p$
\about\ 1.5 if they are optically thin in the wavelength region where $\q
\propto \lambda^{-\beta}$, since $\beta$ is typically 1--2 (Miroshnichenko, Ivezi\'c \& Elitzur 1997).

Our results suggest a natural classification scheme for IR spectra.  For given
grain optical properties, every $\eta$ produces a distinct family of
solutions, with position within the family determined by \tV.  For the most
part, \Tsub\ and \fe\ have only a minor effect.  Therefore, the SEDs of all
astronomical objects that share the same dust density distribution become a
one-parameter family in which each member is fully characterized by its \tV.
This leads to correlations among spectral properties and structure in
color-color diagrams. Scaling breaks down when the dust inner boundary is
controlled by effects other than dust sublimation.  In such sources, the
inner-boundary dust temperature $T_1$ ($< \Tsub$) becomes an additional
independent parameter, varying with the external flux according to $T_1
\propto \Fone^{1/4}$.

Various classes of astronomical objects show evidence for scaling, indicating 
that in these sources the dust inner boundary is controlled by sublimation. A 
common spectrum for compact \HII\ regions was noted in Rowan-Robinson (1979) and 
modelled in RR.  This was further emphasized by Wood and Churchwell (1989), who 
point out that for all ultra-compact \HII\ regions observed by IRAS, ``the 
shapes of flux density distributions are strikingly similar'', a fact also 
recognized by Chini et al.\ (1986).  The trends observed among spectral 
properties of IR radiation from young stellar objects elicited a number of 
classification schemes. Adams, Lada \& Shu (1987) employed the slope of the SED, 
Adams (1990) the visual extinction derived from model spectra, Ladd et al.\ 
(1991) the mean frequency of the spectrum and Myers \& Ladd (1993) the 
corresponding bolometric temperature. Myers \& Ladd also noted that all these 
schemes are equivalent, but the reasons for the trends and for this equivalence 
were not clear. Scaling provides a simple explanation for both.  Furthermore, 
various distributions presented in Myers \& Ladd show that IR spectral 
properties are indeed independent of luminosity, as prescribed by scaling.  
Finally, the structure in color-color diagrams implied by scaling had been 
noticed in the IRAS colors of a variety of objects.  These include T Tau stars 
(Harris, Clegg \& Hughes, 1989), BN objects (Henning, Pfau \& Altenhoff, 1990), 
\HII\ regions (Hughes \& MacLeod, 1989; Wood \& Churchwell, 1989), H$_2$O masers 
in star-forming regions (Wouterloot \& Walmsley, 1986) and late-type stars 
(van der Veen \& Habing, 1988).  Various scaling properties of the latter were noted previously by Rowan-Robinson and Harris (1983 and references therein). 

Dusty winds around late-type stars provide an even tighter form of scaling 
(IE95, IE96).  The structure of these winds is controlled by radiation pressure 
on the dust grains, therefore the density distribution $\eta$ need not be 
prescribed in advance. Instead it is determined from the hydrodynamics coupled 
to the radiative transfer, and the solution does not require any other parameter 
in addition to optical depth.  Therefore, IR emission from all late-type stars 
with a given dust composition can be fully classified in terms of a single 
parameter, \tV.  Detailed analysis shows that the data fully corroborate all the 
correlations predicted by scaling for these sources.  We are in the midst of a 
study that extends this detailed analysis to the IRAS data for all the Galactic 
point sources.  The scaling correlations displayed by these data will enable us 
to identify the dust density distributions of various classes of Galactic 
objects.  The results of this study will be reported separately in a forthcoming 
publication.

\section*{Acknowledgments}

We thank the referee Dr.\ M.\ Rowan-Robinson for his careful reading and useful 
comments which helped improve the manuscript. Support by NSF grant AST--9321847, 
NASA grant NAG 5--3010 and the Center for Computational Sciences of the 
University of Kentucky is gratefully acknowledged.

\appendix

\section                      {GRAIN MIXTURES}

When the dust is composed of different types of grains, its extinction
coefficient is
\eq{
   \k = \rho\sub d\sum_i p_i {\pi a_i^2 \over m_i} Q_{\lambda i}\,.
}
Here $\rho\sub d$ is the overall dust density and $p_i$ is the fraction of that
density due to the $i$-th component ($\sum p_i = 1$). The mass, radius and
extinction efficiency of a single grain from this component are denoted $m_i$,
$a_i$ and $Q_{\lambda i}$, respectively.  The different components reflect
possible variations of chemical composition and/or grain size.  For simplicity
we treat the components as discrete; generalization to the continuous case is
trivial.  With the aid of equation \ref{source}, the dust source function is
\begin{eqnarray}
 \S &= &\sum_i x_{\lambda i}\Big[(1 - \al_i)\B(T\sub{d,i})   \cr
    &&\qquad +\,
    \al_i\int\I(\Omega')g_i(\Omega',\Omega){d\Omega'\over4\upi}\Big],
\end{eqnarray}
where $\al_i$ and $g_i$ are, respectively, the albedo and angular phase
function of the $i$-th component and
\eq{
         x_{\lambda i} = {p_i (\pi a_i^2/m_i) Q_{\lambda i}
             \over \sum_i p_i (\pi a_i^2/m_i) Q_{\lambda i}}
}
is the weight of its contribution to the extinction coefficient at wavelength
$\lambda$.  Each component may have a different temperature, $T_{{\rm d,}i}$,
determined from overall flux conservation.  This equilibrium relation becomes a
vanishing sum of the fractions $p_i$ with coefficients that depend on the
corresponding individual temperatures $T_{{\rm d,}i}$.  Since the temperature
of a dust component cannot depend on its relative abundance, each coefficient
in the sum must vanish separately.  This produces the temperature equation of
the $i$-th component
\eq{Tdi}{
       \int q\sub{a\lambda,i}\B(T\sub{d,i})d\lambda =
       \int q\sub{a\lambda,i}\J d\lambda\,,
}
where $q\sub{a\lambda,i} = q_{\lambda i}(1 - \varpi_{\lambda i})/(1 -
\varpi_{\lambda_0i})$ is the absorption efficiency of the $i$-th component
normalized to the fiducial wavelength $\lambda_0$, analogous to equation
\ref{equilib} for a single-component dust.

A convenient indexing of the dust components is by decreasing sublimation
temperature.  Consider a spherically symmetric configuration with a prescribed
radial variation of overall dust density when the shell inner radius is
controlled by sublimation of the first component at temperature
$T\sub{sub,1}$.  The problem is identical to that of a single-component dust
as long as the temperature of component 2 at dimensionless radius $y$,
determined from equation \ref{Tdi}, remains higher than its sublimation
temperature $T\sub{sub,2}$.  The radius $y_2$ at which component 2 is included
in the mix is thus determined from radiative transfer, and so on.  The only
difference in the solution procedure from the case of single-type dust is that
the dimensionless density profile $\eta(y)$ cannot be fully prescribed
beforehand since the determination of the sublimation radii $y_2$, $y_3$,
etc., requires solution of the radiative transfer problem.  Scaling is
preserved but the exact form of $\eta(y)$ requires an iterative procedure.
The extension of this procedure to arbitrary geometries is straightforward.

\section        {APPROXIMATE SOLUTIONS IN SPHERICAL SYMMETRY}

The radial variation of temperature requires the spatial profile of \J,
equivalent to the spatial variation of both the overall radiative energy
density $J(y) = \int\J(y) d\lambda$ and the frequency profile $\J/J$.  The
former can be obtained from the wavelength-integrated radiative moment
equations when they are closed by some suitable approximation. Auer (1984)
proposed the closure relation
\eq{Auer}{
      3P = 4\upi J + 2F,
}
where $P$ is the bolometric radiation pressure, obtained from the second
angular moment of the intensity.  This relation improves on the classic
Eddington approximation ($3P = 4\upi J$) by providing the correct behavior in
both the optically thin and thick limits.  With this approximation, the
radiation-pressure equation becomes
\eq{
      4\upi{dJ \over dy} = -{F_1\over y^2}\Big[{2\over y} +
               3\tau_0 \qF(y)\eta(y)\Big],
}
where
\eq{
            \qF = \int\q\f d\lambda
}
is the extinction efficiency averaged with the flux spectral shape. Solution of
this equation requires a boundary condition. We assume that the energy density
vanishes when $y \to \infty$, setting aside for a moment the case of finite
shells. Then the solution is
\eq{J1}{
  4\upi J(y) = F_1\Big[{1\over y^2} +
         3\tau_0 \int_y^\infty\!\!\!\!\!\qF(y)\eta(y) {dy\over y^2} \Big].
}
This result requires the flux-averaged \qF, which cannot be calculated without
the spectral shape of the diffuse component of the flux (see equation
\ref{Ffin}).  We will approximate this with the Planck spectral shape at the
local temperature, namely $\Fd \propto \b(T)$.  This can be expected to give an
adequate approximation at large optical depths, less so at small ones, where
the significance of the diffuse component is diminished anyhow.  With this
approximation,
\eqarray{qF}{
      \qF(y) &= &\int\q\fe e^{-\t(y)}d\lambda                 \cr
       &&\qquad +\, \qP(T(y))\Big(1 - \int\fe e^{-\t(y)}d\lambda\Big).
}
This completes the approximation for $J$.  Next we need the spectral shape of
the diffuse component \Jd.  This can be reasonably expected to resemble the
spectral shape of the diffuse flux \Fd, so we approximate it, too, with
$\b(T)$.  Then the radiative equilibrium equation (\ref{equilib}) gives
\eq{J2}{
   4\upi J(y) = 4\sigma T^4 +
   {F_1\over y^2}\int\!\!\fe e^{-\t(y)}\Big(1 - {\qa\over\qaP(T)}\Big)d\lambda.
}
Equations \ref{J1} and \ref{J2} produce the temperature profile as a
function of $y$, with $F_1$, or equivalently $\Psi$ (see equation
\ref{Fscaling}), obtained from the expressions at $y$ = 1.  The results are
\eqarray{approx}{
 \Psi\!\times\!\left(T\over T_1\right)^4 &= &{1\over y^2}
\Big[1-\!\!\int\!\!\fe e^{-\t(y)}\Big(1-{\qa\over\qaP(T)}\Big)d\lambda \Big]
   \cr
   &&\qquad +\, 3\tau_0\int_y^\infty \qF(y)\eta(y) {dy\over y^2},      \cr
 \Psi &= &{\qae\over\qaP(T_1)} + 3\tau_0\int_1^\infty \qF(y)\eta(y)
               {dy\over y^2};
}
note that equation \ref{thin} is recovered when $\tau_0 = 0$. Together with
equation \ref{qF}, this completes the solution for the temperature profile for
any given density profile $\eta$ that extends to infinity.  The solution is
only in implicit form because $T$ enters on the right-hand-side through the
Planck-averages $\qP(T)$ and $\qaP(T)$.  Therefore the complete solution
requires iterations, but these are straightforward and converge rapidly.  We
find that these analytic results provide an excellent approximation for the
exact solutions, typically within 10\%.

When the shell is finite, a boundary condition must be specified for its
energy density at the outer edge \yout.  From the definitions of $J$ and $F$
as angular moments of the intensity it follows that
\eq{
         4\upi J(\yout) = {\gamma F_1\over y^2_{\rm out}}\,,
}
where $\gamma$ varies from 1 when the angular distribution of the emergent
radiation is directional to 2 when it is isotropic.  Leaving $\gamma$ as an
unknown factor, this term is just added to the solution on the right-hand-side
of equation \ref{J1}, simply adding $\gamma/y^2_{\rm out}$ to the
right-hand-side of the two relations in equation \ref{approx}.  So the
analytic approximation for finite shells includes a term that remains
uncertain to within a factor of 2. However, this term is significant only in
shells that are thin both physically and optically ($\yout - 1$, $\tau_0 < 1$)
and our numerical results show that $\gamma \simeq 1.5$ is an adequate
approximation in virtually all cases.

A different approximate solution scheme was proposed in RR.  That scheme does 
not fully incorporate scattering but its iterations include a variable Eddington 
factor, therefore it is in principle capable of convergence to the exact 
solution in the absence of scattering.  Since our scheme involves a prescribed 
Eddington factor (eq.~\ref{Auer}) it only converges to an approximate solution, 
not the exact one. However, when scattering is important it has the edge and we 
usually obtain a level of accuracy comparable to RR with significantly fewer 
computations, mostly because we do not have angular integrations. Furthermore, 
our approximate solution becomes fully explicit in the ``gray opacity" case, the 
approximation that assumes frequency independent extinction. Then \q\ = 1 and 
the solution reverts to the simple, explicit form
\eqarray{gray}{
   \Psi\!\times\!\left(T\over T_1\right)^4 &= &{1\over y^2}
            + 3\tau_0\int_y^\infty \eta(y) {dy\over y^2}, \cr
   \Psi &= &1 + 3\tau_0   \int_1^\infty \eta(y) {dy\over y^2}.
}
In the case of a finite shell, the right-hand-sides of both equations again
are augmented by $\gamma/y^2_{\rm out}$. This gray-opacity result follows
directly from the Auer closure relation and does not require assumptions about
the spectral shape of the diffuse component.  It provides a closed-form
expression that readily explains many properties of the full solutions,
as pointed out in section 5.

\section                   {NUMERICAL PROCEDURES}

The code DUSTY, which solves the spherical radiative transfer problem employing 
the scaling approach described here, is publicly available for general use 
(Ivezi\'c, Nenkova \& Elitzur 1997).  The numerical calculations are  
conveniently performed with the aid of the dimensionless energy density
\eq{
                  \u = {4\upi y^2 \over \Fone} \J.
}
Extracting the $y^2$ radial dilution reduces the dynamic range of the
density, helping the numerical accuracy.  For isotropic scattering, \u\ at
radius $y$ obeys the integral equation
\eqarray{u}{
      \u(y) &= &\fe e^{-\t(y)}               \cr
       && +\, {1\over2} \int\!\!
   \left[\al\u(y') +
(1 - \al)\Psi\left({T'\over T_1}\right)^{\!\!4}\b(T') \right] \cr
       && \times e^{\t(y',\mu) - \t(y,\mu)}
    \left({y\over y'}\right)^{\!\!2} d\t(y',\mu)\, d\mu\, ,
}
where $\mu = \cos\theta$. In this equation $T' = T(y')$ is obtained from
\eq{
         {\qaP(T')\over\qaP(T_1)}\left({T'\over T_1}\right)^4 =
         {\int\qa\u(y')d\lambda\over\int\qa\u(1)d\lambda}
}
and $\Psi$ from
\eq{
               \Psi =  {1\over\qaP(T_1)}\int\qa\u(1)d\lambda.
}
Once a radial grid is prescribed, the numerical evaluation of the integral of
any radial function is transformed into multiplication with a matrix of weight
factors determined purely by the geometry (cf Schmid-Burgk, 1975).  Therefore,
if $T(y)$ and $\Psi$ are given, \u\ can be obtained for all radial grid points
through matrix inversion.  This provides a direct solution to the full
scattering problem, obviating the need to iterate over $\u(y)$ itself.

We start the calculation with $T(y)$ and $\Psi$ from the analytic solution of
appendix B, calculate the corresponding $\u(y)$ and iterate until convergence.
Because the scattering problem is solved directly by the matrix inversion,
convergence is very rapid; an accuracy of $10^{-4}$ is achieved with fewer
than 10 iterations even for the largest optical depths considered  here.  The
solution is accomplished once the density \u\ does not vary, within a
prescribed tolerance, when the grid density is increased.  The number of
radial grid points we needed varied from \about\ 20 for moderate optical
depths ($\tV \la 10$) to 80 for \tV\ = 1000. The number of angular grid points
used in the $\mu$-integration is typically 2--3 times the number of radial
grid points.  Once the energy density \u\ is found, the flux spectral shape
$\f(y)$ is calculated from the same integral with $d\mu$ replaced by $\mu
d\mu$.  

We have compared our code with a number of benchmark solutions and verified its 
accuracy to better than 0.1\% (Ivezi\'c et al. 1997). Most of the computer time 
is spent on calculation of the weight matrix for spatial integrations.  This 
matrix is determined purely by geometry and does not vary between iterations, 
giving our method an edge in calculation speed over other schemes (e.g.\ 
iterations over variable Eddington factor).  In particular, large optical depths 
do not increase the computational difficulties as much as they do in purely 
iterative schemes; we encountered no difficulties with \tV\ as large as 2000. In 
addition, the inclusion of scattering in the direct matrix inversion eliminates 
the problems that large albedos present in other schemes.  Our method remains 
stable even for albedos that approach unity at some wavelengths.

\label{lastpage}

\end{document}